\documentclass[twocolumn,%showpacs,preprintnumbers,
 superscriptaddress,amsmath,amssymb,nofootinbib]{revtex4}
\usepackage{amsfonts}
\usepackage{amsmath}
\usepackage{bbm}
\usepackage{amssymb}
\usepackage{graphicx}%
\usepackage{footmisc}
\usepackage{bm}
\usepackage{braket}
\usepackage{CJK}
\usepackage{hyperref}
\usepackage{bbm}
\usepackage{cleveref}
\usepackage{longtable,booktabs}

\crefname{equation}{Eq.}{Eqs.}

\newcommand{\norm}[1]{\lVert#1\rVert}
\newcommand{\Tr}{\text{Tr}}

\begin{document}
\begin{CJK*}{UTF8}{gbsn}
\title{Fluctuation Theorems for Multitime Processes}

\author{Zhiqiang Huang (黄志强)}
%\thanks{}hzq@wipm.ac.cn
\email{hzq@wipm.ac.cn}
\affiliation{Innovation Academy for Precision Measurement Science and Technology\footnote{The State Key Laboratory of Magnetic Resonance and Atomic and Molecular Physics, the National Centre for Magnetic Resonance in Wuhan, and the Wuhan Institute of Physics and Mathematics.}, CAS, Wuhan 430071, China}

\date{\today}

\begin{abstract}
    In this paper, we extend the fluctuation theorems used for quantum channels to multitime processes. The fluctuation theorems for quantum channels are less restrictive. We show that the given entropy production can be equal to the result of a closed system environment. The assumption that the system evolves under a completely positive and trace preserving map is quite general, but it is more specific for cases in which the system is initially correlated with the environment. System-environment correlations arise naturally in multitime processes, with which we can give clear and physical interpretations regarding the effects of correlations. Multitime processes can provide many-body channels. The Choi state of such a many-body channel is called a process tensor. One can derive channels by executing the process tensor on a set of operations. We establish a general quantum fluctuation theorem framework for a many-body channel and its 
derived channels. In this framework, the effects of correlations are reflected in a Markovian property. For Markovian processes, we can extend the two-point measurement to a three-point measurement and obtain that the fluctuation theorems contain complete information about the intermediate state. For non-Markovian processes, the complete measurement of the intermediate state leads to conflicts. Therefore, we use a general measurement, which only provides partial information, for the intermediate state. The corresponding fluctuation theorems show that memory effects can reduce these fluctuations. This is consistent with the fact that system states can be recovered under non-Markovian processes.
%This work may be also helpful in understanding and giving Markov order.
\end{abstract}

%\pacs{03.65.Yz}
%\keywords{Lieb-Robinson bound}

\maketitle

\section{Introduction}\label{INTRO}
The fluctuation theorem (FT) deals with the relative probability that some extensive quantities of a system which is currently away from thermodynamic equilibrium will increase or decrease over a given amount of time. For non-equilibrium statistical mechanics, the ergodic hypothesis usually does not apply. Starting from the deterministic equation such as Liouville's equation or the von Neumann equation is a safer way. However,  the exact solutions of these equations are very difficult to obtain. The FT is a suitable method for non-equilibrium statistical mechanics. The proof of  FT is based on the deterministic equation. With some simple assumptions, it  can describe some universal properties of nonequilibrium fluctuations such as irreversible work fluctuations and entropy fluctuations.  It also gives a generalization of the second law of thermodynamics. The FT bridges the microscopic dynamics and the macroscopic observations. Hence, FT is of fundamental importance to non-equilibrium statistical mechanics. 

Research on fluctuation theorems (FTs) for closed quantum systems has been fruitful in recent years \cite{EHM09, LP21}. 
FTs reflect the symmetry between forward evolution and backward evolution processes. In a closed quantum system, the forward evolution process is denoted as $U$, which is unitary. The backward evolution process is the time-reverse evolution process $U^\dagger$, which is also unitary. Under backward evolution, all the initial states can be recovered. In open quantum systems, these properties are not available. For example, when a system and its environment are initially uncorrelated, the dynamical map of the open system is completely positive and trace preserving (CPTP). The volume of the quantum state space of the system can shrink under the dynamical map \cite{LP13}. The backward map cannot simultaneously guarantee exact recovery and the CPTP properties. A recent work \cite{KK19} chose the Petz recovery map as a backward map and proved a general FT for quantum channels. 
Although the evolutions of open systems are derived from closed system environments, their FTs look very different. The average entropy production of a closed system environment is equal to the quantum relative entropy between the final density matrix of the forward evolution process and the initial density matrix of the backward evolution process. The average entropy production of the open system is equal to the decrease in the quantum relative entropy between the initial state and reference state. These seemingly different results are actually closely related. It has been proven that they are equal when the global unitary operation $U$ satisfies the strict energy conservation condition \cite{ LP21}.

When memory effects are present, the associated dynamical map cannot be CPTP \cite{BLPV16}. The FTs for quantum channels do not seem to apply to such circumstances. Moreover, in the FTs for quantum channels, irrecoverability is the key to the fluctuations. The distinguishability of the states should not increase in the FTs for quantum channels. This is in conflict with the fact that the states can be recovered in non-Markovian processes \cite{BLP09}. Furthermore, the average entropy production is nonnegative in the FTs for quantum channels. This means that the corresponding physical observation monotonically increases during evolution procedure, and this is also in conflict with a non-Markovian process. These contradictions suggest that the common FTs derived from two-point measurements (TPMs) are not applicable to non-Markovian processes. To resolve this issue and obtain the FTs for non-Markovian processes, one approach is to consider the entire system of interest, including its environment. 

Here, we propose a different approach inspired by the process tensor \cite{M12,PRFPM18b}. In the framework of the process tensor, the quantum system is undergoing a process that one can split into arbitrary discrete time steps. There multitime processes fully characterize the interaction between system and environment. The process tensor is separated from the Choi representation for multitime processes. This tensor is the Choi state of a many-body channel, which contains all the available evolutionary information. The many-body channel itself is a CPTP map regardless of whether the evolution is Markovian. 
One can trace a multitime process back by executing the process tensor on identity operations. For more general CPTP operators, we obtain other multitime channels. By applying the FTs developed for quantum channels, it is easy to obtain the FTs for many-body channels and derived channels.

In this paper, we study the FTs for many-body channels and derived channels. In \cref{FTSP}, we first rewrite the proof of the FTs for quantum channels with superoperators and then with Choi representation. This helps to simplify the proof. We also find an equivalence relation between the FTs for closed systems and the FTs for open systems without strict energy conservation conditions. After these preparations, we study the FTs for multitime processes in \cref{MPC}. We first present a many-body channel and several derived channels that are closely related to multipoint measurements. We show that the Petz recovery map of the many-body channel cannot derive channels as we do for the many-body channel. Only for a Markovian process can the Petz recovery map be time-ordered and linkable. Consistently, one can trace the backward evolution process
from the Petz recovery map of the many-body channel for a Markovian process. In such a form, the intermediate state of the system is completely measurable \cite{DPS04}. For non-Markovian processes, we can only trace the backward evolution process from the Petz recovery map of the derived channel. We mainly focus on a derived channel that is closely related to a general measurement \cite{F95} of the intermediate state. We prove the corresponding FTs and show how the memory effects reduce the system fluctuations.

\section{Preliminaries}\label{FTSP}
\subsection{Proving FTs with a superoperator}\label{PFTWS}
We reformulate the FTs for quantum channels \cite{KK19} with a superoperator \cite{P15}. We regard an operator $O$ as a state $|O)$. The inner product of the operators $(O_A|O_B):=\Tr(O_A^\dagger O_B)=(O_B|O_A)^*$. The operator vector space is orthonormalized as follows:
\begin{equation}
    (\Pi_{kl}|\Pi_{ij})=\delta_{ik}\delta_{jl},
\end{equation}
where $\Pi_{ij}=\ket{i}\bra{j}$. The completeness relation is
\begin{equation}
    \hat{I}=\sum_{ij}|\Pi_{ij})(\Pi_{ij}|.
\end{equation}
The quantum channel $\mathcal{N}$ is a superoperator that maps a density matrix $|\rho)$ to another density matrix $|\rho')=\mathcal{N}|\rho)$. 

The forward transition matrices are defined as
\begin{equation}\label{FTP}
    T_{{ij}\to{k'l'}}=(\Pi_{k'l'}|\mathcal{N}|\Pi_{ij}),
\end{equation}
where quantum channel $\mathcal{N}(\cdot)=\sum_i M_i(\cdot)M_i^\dagger$ is a CPTP map. The forward transition matrices contain all evolution information. It is easy to obtain the transition probabilities from this information. By using the definition of the inner product of operators, we have 
\begin{equation}\label{FTP2}
    T_{{ij}\to{k'l'}}=(\Pi_{ij}|\mathcal{N}^\dagger|\Pi_{k'l'})^*.
\end{equation}
The superoperator $\mathcal{N}^\dagger(\cdot)=\sum_i M_i^\dagger (\cdot)M_i$ is not a CPTP map, so it cannot be used as a backward map. The Petz recovery map \cite{P86}
 \begin{equation}
    \mathcal{R}_{\gamma}=\mathcal{J}_\gamma^{1/2}\circ\mathcal{N}^\dagger\circ\mathcal{J}_{\mathcal{N}(\gamma)}^{-1/2}
\end{equation}
is commonly used as a backward map. $\gamma$ is called the reference state, and $\mathcal{J}^\alpha_O(\cdot):=O^\alpha(\cdot){O^\alpha}^\dagger$ is defined as a rescaling map. The Petz recovery map is CPTP. It can fully recover the reference state $ \mathcal{R}_{\gamma}\circ \mathcal{N}(\gamma)=\gamma$ but generally cannot recover other states.
The trace preserving property of the Petz recovery map is easily obtained from
\begin{equation}
    (I|\mathcal{J}_\gamma^{1/2}\circ\mathcal{N}^\dagger\circ\mathcal{J}_{\mathcal{N}(\gamma)}^{-1/2}|O) =(O|\mathcal{J}_{\mathcal{N}(\gamma)}^{-1/2}\circ\mathcal{N}|\gamma)^*= \Tr(O).
\end{equation}
With the Petz recovery map, \cref{FTP2} becomes
\begin{equation}\label{RBFR}
    T_{{ij}\to{k'l'}}=(\mathcal{J}_\gamma^{-1/2}\Pi_{ij}| \mathcal{R}_{\gamma}|\mathcal{J}_{\mathcal{N}(\gamma)}^{1/2}\Pi_{k'l'})^*.
\end{equation}
The operators obtained from the rescaling map are no longer normalized. After normalization, we have so-called reference-rescaled operators \cite{EDRV15}:
\begin{equation}\label{FOSPO}
    |\Pi'_{k'l'})= \frac{1}{Z'_{k'l'}} |\mathcal{J}_{\mathcal{N}(\gamma)}^{1/2}\Pi_{k'l'}),
\end{equation}
where the factor
\begin{align}
    Z'_{k'l'}=\sqrt{ (\mathcal{J}_{\mathcal{N}(\gamma)}^{1/2}\Pi_{k'l'}|\mathcal{J}_{\mathcal{N}(\gamma)}^{1/2}\Pi_{k'l'})}\notag\\
    =\norm{\mathcal{J}_{\mathcal{N}(\gamma)}^{1/2}\Pi_{k'l'}}_2=\sqrt{ (\Pi_{k'}|\mathcal{N}(\gamma))(\mathcal{N}(\gamma)|\Pi_{l'})}.
\end{align}
Similarly, we have $ |\Pi'_{ij})= |\mathcal{J}_\gamma^{-1/2}\Pi_{ij})/Z_{ij}$ and
\begin{align}
    Z_{ij}=\sqrt{ (\mathcal{J}_{\gamma}^{-1/2}\Pi_{ij}|\mathcal{J}_{\gamma}^{-1/2}\Pi_{ij})}\notag\\
    =\norm{\mathcal{J}_{\gamma}^{-1/2}\Pi_{ij}}_2=\sqrt{ (\Pi_{i}|\gamma^{-1})(\gamma^{-1}|\Pi_{j})}.
\end{align}
With reference-rescaled operations, \cref{RBFR} can be rewritten as
\begin{equation}\label{RBFRrro}
    T_{{ij}\to{k'l'}}=\tilde{T}_{{ij}\leftarrow{k'l'}}^* Z_{ij}Z'_{k'l'},
\end{equation}
where
\begin{equation}
    \tilde{T}_{{ij}\leftarrow{k'l'}}=(\Pi'_{ij}|\mathcal{R}_{\gamma}
    |\Pi'_{k'l'})
\end{equation}
denotes the backward transition matrices. \cref{RBFRrro} shows a clear relationship between the forward transition matrices and the backward transition matrices.

Suppose that the system evolves from $\rho=\sum_u p_u \ket{\psi_u}\bra{\psi_u}$ to $\rho'=\mathcal{N}(\rho)=\sum_{v'} {p'}_{v'} \ket{\phi'_{v'}}\bra{\phi'_{v'}}$. Then, the two-point measurement (TPM) quasiprobability distribution for the forward process is
\begin{equation}\label{FTPT}
    P^{u, v'}_{ij, k'l'}=p_u (\Pi_{\phi'_{v'}}  |\Pi_{k'l'}) (\Pi_{k'l'}|\mathcal{N}|\Pi_{ij})(\Pi_{ij}|\Pi_{\psi_u}),
\end{equation}
where $\Pi_{\psi_u}= \ket{\psi_u}\bra{\psi_u}$ and $\Pi_{\phi'_{v'}}=\ket{\phi'_{v'}}\bra{\phi'_{v'}}$.
It is easy to show that 
\begin{align}\label{HET}
   ( P^{u, v'}_{ij, k'l'})^*=p_u (\Pi_{\phi'_{v'}}  |\Pi_{k'l'}^\dagger) (\Pi_{k'l'}^\dagger|\mathcal{N}|\Pi_{ij}^\dagger)(\Pi_{ij}^\dagger|\Pi_{\psi_u})\notag\\
   = P^{u, v'}_{ji, l'k'}.
\end{align}
The TPM quasiprobability distribution for the backward process is
\begin{equation}
    {P'}^{u, v'}_{ij, k'l'}=p_{v'}(\Pi_{\psi_u}|\Pi'_{ij}) (\Pi'_{ij}|\mathcal{R}_{\gamma}|\Pi'_{k'l'}) (\Pi'_{k'l'}  |\Pi_{\phi'_{v'}}).
\end{equation}
The entropy production can be defined as
\begin{equation}
    \sigma^{u\to v'}_{ij\to k'l'}=\delta s^{u\to v'}-\delta q_{ij\to k'l'}
\end{equation}
where $\delta s^{u\to v'}=-\log( {p'}_{v'})+\log(p_u)$ and $\delta q_{ij\to k'l'}=-\log(Z_{ij}Z'_{k'l'})$.
According to its definition, the entropy production satisfies
\begin{equation}\label{TIS}
    \sigma^{u\to v'}_{ij\to k'l'}= \sigma^{u\to v'}_{ji\to l'k'}
\end{equation}
The entropy production distribution can be obtained directly from the TPM quasiprobability as follows: 
\begin{equation}\label{DEP}
    P_\to(\sigma)=\sum_{u,i,j}\sum_{v',k',l'} P^{u, v'}_{ij, k'l'}\delta(\sigma-\sigma^{u\to v'}_{ij\to k'l'}).
\end{equation}
The entropy production distribution under the recovery map can be defined as
\begin{equation}\label{DEPR}
    P_\leftarrow(\sigma)=\sum_{u,i,j}\sum_{v',k',l'} {P'}^{u, v'}_{ij, k'l'}\delta(\sigma+\sigma^{u\to v'}_{ij\to k'l'}).
\end{equation}
The basis $\{\ket{i}\}$ is chosen such that it diagonalizes the reference state $\gamma$, and $\{\ket{k'}\}$ is chosen as the eigenbasis of $\mathcal{N}(\gamma)$. These facts lead to $|\Pi'_{ij})=|\Pi_{ij})$ and $|\Pi'_{k'l'})=|\Pi_{k'l'})$. Combining this with \cref{DEP,DEPR,HET,RBFRrro,TIS}, we derive the following relation:
\begin{equation}\label{FTR}
    \frac{P_\to(\sigma)}{P_\leftarrow(-\sigma)}=e^{\sigma}.
\end{equation}
The FT $\braket{ e^{-\sigma} }=1$ can be easily obtained from this relation, and the generalized second law yields
\begin{equation}\label{FOAV}
    \braket{\sigma}=S(\rho||\gamma)-S(\mathcal{N}(\rho)||\mathcal{N}(\gamma))\geq 0,
\end{equation}
which is consistent with the conclusion that the distinguishability of quantum states does not increase under a CPTP map.

\subsection{Relationship between two types of FTs}
Here, we show the relationship between the FTs for quantum channels and the FTs for closed systems. The premise is that their evolution processes must be consistent; that is, the dynamical map of an open system is
\begin{equation}\label{DMOOS}
    \mathcal{N}(\cdot)=\Tr_E  [{U}_{SE} (\cdot\otimes \rho_E^0){U}_{SE}^\dagger].
\end{equation}
Under a strict energy conservation condition and by choosing $\tilde{\rho}_{SE}=\rho'_S\otimes \rho_E$, the entropy production of a system environment is \cite{LP21}
\begin{equation}
    \Sigma=S(\rho_S||\rho_S^{\text{th}})-S(\rho'_S||\rho_S^{\text{th}}).
\end{equation}
$\Sigma$ is equal to the entropy production $\braket{\sigma}$ yielded when setting the global fixed points $\rho_S^{\text{th}}$ as the reference state. 

The strict energy conservation condition is not necessary to bridge two FTs. The reference state does not have to be the global fixed points. 
For an arbitrary reference state, the entropy production of \cref{FOAV} becomes
\begin{align}\label{EPOQC}
    \braket{\sigma}=S(\rho_S\otimes \rho_E^0||\gamma_S\otimes \rho_E^0)-\Tr_{SE}(\rho'_{SE}(\ln \rho'_S -\ln \gamma'_S))\notag\\
    =\Tr_{SE}(\rho'_{SE}(\ln \rho'_{SE} -(\ln \rho'_S +\ln \gamma'_{SE} -\ln \gamma'_S)))\notag \\
    =S(\rho'_{SE}||\tilde{\rho}_{SE})
\end{align}
where $\tilde{\rho}_{SE}=\exp(\ln \rho'_S +\ln \gamma'_{SE} -\ln \gamma'_S)$. We use \cref{DMOOS} in the equality.

Here, we briefly discuss the meaning of $\tilde{\rho}_{SE}$. If the dynamical map does not change the relative entropy between $\rho$ and $\gamma$, then we have
\begin{equation}\label{REDC}
    \braket{\sigma}=0=S(\rho'_{SE}||\gamma'_{SE})-S(\rho'_S||\gamma'_S).
\end{equation}
In the sandwiched R\'enyi divergence, the equivalent condition of \cref{REDC} is \cite{L16}
\begin{align}\label{ECOREDC}
    {\gamma'_S}^\beta ({\gamma'_S}^\beta \rho'_S{\gamma'_S}^\beta)^{\alpha-1} {\gamma'_S}^\beta \otimes I_{E}\notag\\
   = {\gamma'_{SE}}^\beta ({\gamma'_{SE}}^\beta \rho'_{SE}{\gamma'_{SE}}^\beta)^{\alpha-1} {\gamma'_{SE}}^\beta,
\end{align}
where $\beta=(1-\alpha)/(2\alpha)$. Condition (\ref{ECOREDC}) can be rewritten as
\begin{equation}
    \rho'_{SE}=\mathcal{R}_{S\to SE}^{\alpha,\gamma,\mathcal{N}}(\rho'_S):=\mathcal{J}_{\gamma'_{SE}}^{-\beta}[\mathcal{J}_{\gamma'_{SE}}^{-\beta}\circ\mathcal{J}_{\gamma'_S}^\beta (\mathcal{J}_{\gamma'_S}^\beta \rho'_S)^{\alpha-1}]^{\frac{1}{\alpha-1}}.
\end{equation}
It is easy to show that 
\begin{equation}\label{RCMTFT}
    \lim_{\alpha\to 1} \mathcal{R}_{S\to SE}^{\alpha,\gamma,\mathcal{N}}(\rho'_S)=\tilde{\rho}_{SE}.
 \end{equation}
$\tilde{\rho}_{SE}$ is given by the map $\mathcal{R}_{S\to SE}^{\alpha,\gamma,\mathcal{N}}$ and is very different from the normal chosen $\tilde{\rho}_{SE}=\mathcal{R}_{S\to SE}(\rho'_{S})=\rho'_{S}\otimes\rho_{E}$. The map is also different from the maps listed in \cite{LP21}. System-environment correlations are allowed in $\mathcal{R}_{S\to SE}^{\alpha,\gamma,\mathcal{N}}(\rho'_S)$. To our limited knowledge, the map $\mathcal{R}_{S\to SE}^{\alpha,\gamma,\mathcal{N}}$ is new and needs further research.

When we choose $\tilde{\rho}_{SE}=\mathcal{R}_{S\to SE}^{\alpha,\gamma,\mathcal{N}}(\rho'_S)$ as the initial state of the backward process, the given entropy production of the entire system, including the environment, is 
\begin{equation}
    \Sigma=S(\rho'_{SE}||\tilde{\rho}_{SE})
\end{equation}
according to the FTs for closed quantum systems \cite{EHM09}.
It is equal to the entropy production given in \cref{EPOQC}. Hence, the two FTs are consistent with each other.

One advantage of the FTs for quantum channels is that the entropy production $\braket{\sigma}$ is also closely related to the TPM. If the reference state can be described with intensive properties and extensive properties, such as 
\begin{equation}\label{RSIPEP}
    \gamma(t)=\exp(-\sum_i \beta_i O^i(t))/Z(t),
\end{equation}
then we have that $\ln \gamma=-\sum_i \beta_i  O^i(0) +\beta F(0)$ and $\ln \mathcal{N}(\gamma)=-\sum_i \beta_i  O^i(t) +\beta F(t)$. Under these circumstances, 
the second law given by \cref{FOAV} is related to the observables
\begin{equation}
    \beta \braket{w }-\Delta S\leq \beta \Delta F,
\end{equation}
where $ \Delta S=S(\rho')-S(\rho)$ and 
\begin{equation}
    \beta\braket{w}=\sum_i  \beta_i [ \Tr O^i(t)\rho'_S-\Tr O^i(0)\rho_S].
\end{equation}
Any map involving a thermal environment necessarily has its thermal state as a fixed point when strict energy conservation holds \cite{LP21,BP07}, so the above assumption (\ref{RSIPEP}) is fairly general.

Another advantage of the FTs for quantum channels is that the initial state of the system is arbitrary. We do not need to suppose that the initial state of the system is in equilibrium, while this assumption is often used in the FTs for closed systems.

\subsection{Rewriting FTs with the Choi-Jamio\l{}kowski isomorphism}\label{CJFT}
We rewrite the FT proof with the Choi-Jamio\l{}kowski isomorphism. According to the proof in \cref{PFTWS}, the TPM and quantum channel are the keys to the FTs. The Choi-Jamio\l{}kowski isomorphism can convert the quantum channel to a Choi state \cite{L06} and realize the TPM with one operator. This helps to generalize the FTs for quantum channels to multitime processes. 

In the Choi-Jamio\l{}kowski isomorphism, one must introduce an auxiliary system $A$ with the same dimensionality as the original system. The density matrix of the maximally entangled state between the ancilla and Choi-Jamio\l{}kowski system is as follows:\begin{equation}
    \Phi^{AS}=1/N\sum_{ij} \Pi_{ij}^A\otimes  \Pi_{ij}^S,
\end{equation}
where $N$ is the Hilbert space dimensionality of the system.
The density operator $I^A\otimes \mathcal{N}^S|\Phi^{AS})$ is called the Choi state. The maximally entangled state has the following property:
\begin{equation}
    N (\rho_A|\Phi^{AS})={\rho_S}=|\rho_S),
\end{equation}
with which we can rewrite the forward transition matrices in \cref{FTP} as
\begin{equation}\label{FTPA}
    T_{{ij}\to{k'l'}}=N ( \Pi_{ij}^A\otimes\Pi_{k'l'}^S|\mathcal{N}_{AS}|\Phi^{AS}),
\end{equation}
where $\mathcal{N}^{AS}= I^A\otimes\mathcal{N}^S$ is still a CPTP map. In this form, the TPM is realized with the operator $N( \Pi_{ij}^A\otimes\Pi_{k'l'}^S|$. The forward transition matrices can also be treated as the inner product of the measurement operator and the Choi state. The relation in \cref{RBFR} becomes
\begin{align}\label{POCPTP}
    T_{{ij}\to{k'l'}}=N((\mathcal{J}_\gamma^S)^{-1/2} \Phi^{AS}|\notag\\
    \mathcal{R}_{\gamma}^S \otimes I^A| (\mathcal{J}_{\mathcal{N}(\gamma)}^S)^{1/2}\Pi_{ij}^A\otimes\Pi_{k'l'}^S)^*.
\end{align}
For arbitrary operators $O^1$ and $ O^2$ and a superoperator $\mathcal{M}$, the maximally entangled state $\Phi^{AS}$ can switch the rescaling map:
\begin{align}
    (O_A^1\otimes O_S^2|\mathcal{M}_S|\mathcal{J}_{\rho_S}\Phi^{AS})= ( O_S^2|\mathcal{M}_S|\mathcal{J}_{\rho_S}O_S^1)\notag\\
    =((\mathcal{J}_{\rho_A} O_A^1)\otimes O_S^2|\mathcal{M}_S|\Phi^{AS}).
\end{align}
This allows us to move the full rescaling map in \cref{POCPTP} to the same side and obtain
\begin{align}\label{jgn}
    T_{{ij}\to{k'l'}}=N( \Phi^{AS}|\notag\\
    \mathcal{R}_{\gamma}^S \otimes I^A| (\mathcal{J}_{\gamma,\mathcal{N}}^{AS})^{1/2}\Pi_{ij}^A\otimes\Pi_{k'l'}^S)^*,
\end{align}
where $\mathcal{J}_{\gamma,\mathcal{N}}^{AS}=(\mathcal{J}_\gamma^A)^{-1}\otimes \mathcal{J}_{\mathcal{N}(\gamma)}^S$.
The backward transition matrices can be defined as 
\begin{equation}
    \tilde{T}_{{ij}\leftarrow{k'l'}}=N(\Phi^{AS}|  \mathcal{R}_{\gamma}^S \otimes I^A| {\Pi'}_{ij}^A\otimes{\Pi'}_{k'l'}^S).
\end{equation}
where
\begin{equation}
    | {\Pi'}_{ij}^A\otimes{\Pi'}_{k'l'}^S)= \frac{1}{Z_{ij}Z'_{k'l'}} | (\mathcal{J}_{\gamma,\mathcal{N}}^{AS})^{1/2}\Pi_{ij}^A\otimes\Pi_{k'l'}^S).
\end{equation}
 \cref{RBFRrro} still holds, and the TPM quasiprobability distribution for the forward process is now written as
\begin{equation} \label{TPMQDF}
    P^{u, v'}_{ij, k'l'}=p_u  N(   \Pi_{\psi_u}^A\otimes \Pi_{\phi'_{v'}}^S|\Pi_{ij}^A\otimes\Pi_{k'l'}^S)( \Pi_{ij}^A\otimes\Pi_{k'l'}^S|\mathcal{N}^{AS}|\Phi^{AS}).
\end{equation}
The TPM quasiprobability distribution for the backward process becomes
\begin{align}
    P'^{u, v'}_{ij, k'l'}=p'_{v'}  N( \Phi^{AS}|\mathcal{R}^{AS}|{\Pi'}_{ij}^A\otimes{\Pi'}_{k'l'}^S) \notag\\
    \times(  {\Pi'}_{ij}^A\otimes{\Pi'}_{k'l'}^S|{\Pi}_{\psi_u}^A\otimes {\Pi}_{\phi'_{v'}}^S).
\end{align}
The other proofs and results are not different from those in \cref{PFTWS}.

\section{Multitime processes}\label{MPC}
\begin{figure*}[htb]
    \centering
    \includegraphics[width=0.9\textwidth]{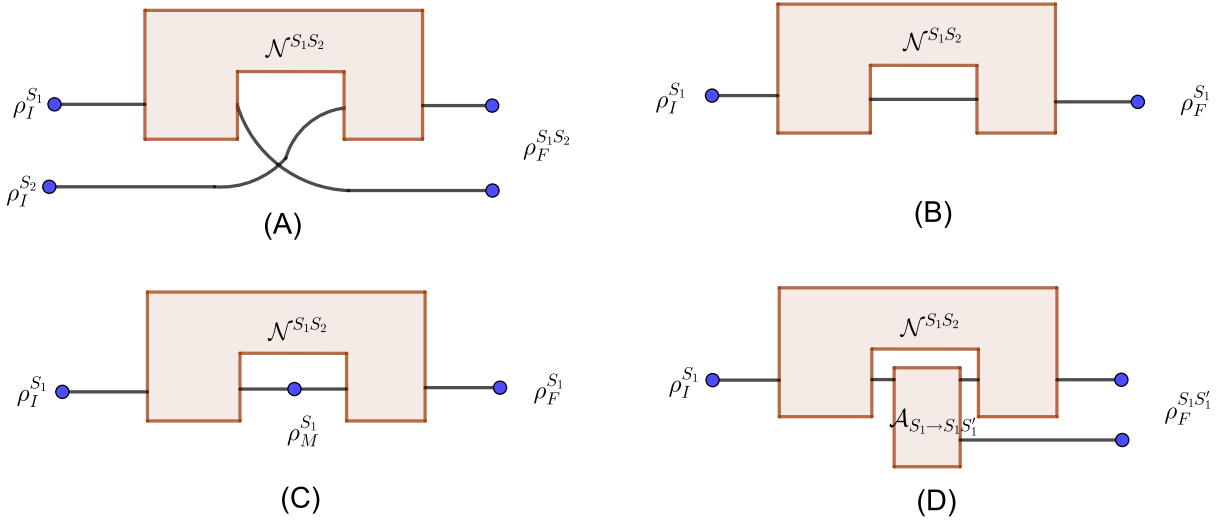}
    \caption{The many-body channel (A) can be extracted from multitime processes. It is time-ordered and linkable. The many-body channel itself can be treated as multitime processes that freshly prepare the system state at each step. When linking the steps without measurements, channel (B) emerges. For a Markovian process, the measurements over the intermediate state do not lead to any conflict. Therefore, we can link the steps with the measurements and obtain (C). For a non-Markovian process, we can only insert operations and discuss the FTs with derived channel (D).
    }\label{LINK}
    \end{figure*}

As in the procedure of the process tensor \cite{M12,PRFPM18b}, we assume that the system and environment are initially uncorrelated. After the first step, the unrestricted unitary evolution of the system environment enables the creation of correlations. Such multitime processes can map the initial system $|\rho_S)$ to
\begin{equation}
    |\rho'_S)=(I_E| \mathcal{U}_{SE}^n\circ \ldots\circ \mathcal{U}_{SE}^1|\rho_S\otimes \rho^0_E),
\end{equation}
where $\mathcal{U}_{SE}(\cdot)={U}_{SE} (\cdot){U}_{SE}^\dagger$ and $\rho^0_E$ is the initial density matrix of the environment. The corresponding process tensor is
\begin{equation}\label{NSEC}
    |\mathcal{T}_n)=\Tr_E  \mathcal{U}_{S^n E}^n\circ \ldots\circ \mathcal{U}_{S^1 E}^1\circ\mathcal{A}| (\prod_{i=1}^n  \Phi^{A^iS^i})), 
\end{equation}
where $\mathcal{A}(\cdot)=(\cdot)\otimes \rho^0_E$. This process tensor is the Choi state of the following many-body channel: 
\begin{equation}
    \mathcal{N}^{S^1\dots S^n}= \Tr_E  \mathcal{U}_{S^n E}^n\circ \ldots\circ \mathcal{U}_{S^1 E}^1\circ\mathcal{A},
\end{equation}
which can also be treated as the process by which multiple systems interact with a single environment in turn. If the evolution is a Markovian process, the many-body channel can be decomposed into several CPTP maps \cite{PRFPM18}:
\begin{equation}\label{MMBC}
    \mathcal{N}^{S^1\dots S^n}= \mathcal{N}^{S^n}_n \circ \dots \circ  \mathcal{N}^{S^1}_1.
\end{equation}
Such a decomposition does not hold when the evolution process is non-Markovian. The output state of the Markovian many-body channel in \cref{MMBC} is a tensor product state when the input state is a tensor product state:
\begin{equation}
    |\rho^{S^1\dots S^n}_F)=N^n(\rho^{A^1}_I\otimes \dots \otimes\rho^{A^n}_I|\mathcal{T}_n)=    |\rho^{S^1}_F\otimes \dots \rho^{S^n}_F).
\end{equation}
The decomposition also does not hold for non-Markovian evolution.

The many-body channel in \cref{NSEC} is \emph{time-ordered}, which means that the future process will not affect the current state:
\begin{align}
    (I^{S^{i+1}\dots S^{n}}|\mathcal{N}^{S^1\dots S^n} |\rho_1\otimes \dots \otimes \rho_n)\notag \\
    =\mathcal{N}^{S^1\dots S^{i}}|\rho_1\otimes \dots \otimes\rho_i)\times (I^{S^{i+1}\dots S^{n}}| \rho_{i+1}\otimes\dots \otimes  \rho_n).
\end{align}
However, the current state can affect the future evolution:
\begin{align}
    (I^{S^{1}\dots S^{i}}|\mathcal{N}^{S^1\dots S^n} |\rho_1\otimes \dots \otimes \rho_n)\notag \\
    =\mathcal{N}_{\rho_{1}\otimes\dots \otimes  \rho_i}^{S^{i+1}\dots S^{n}} |\rho_{i+1}\otimes \dots \otimes \rho_n),
\end{align}
where the maps $\mathcal{N}_{\rho_{1}\otimes\dots \otimes  \rho_i}^{S^{i+1}\dots S^{n}}$ vary with the historical states $\rho_{1}\otimes\dots \otimes  \rho_i$. Only when the evolution process is Markovian can the future evolution be history-independent.
 The many-body channel in \cref{NSEC} is also \emph{linkable}; that is, a new quantum channel can be derived by linking the output state of the previous step with the input state of the next step:
 \begin{equation}
    \mathcal{N}^{S^1\dots S^i S^{i+2}\dots  S^n}= \Tr_E  \mathcal{U}_{S^n E}^n\circ\ldots\circ( \mathcal{U}_{S^{i} E}^{i+1} \mathcal{U}_{S^{i} E}^{i} )\ldots\circ \mathcal{U}_{S^1 E}^1\circ\mathcal{A}.
\end{equation}
The multitime evolution can be obtained by linking all steps:
\begin{equation}\label{EWIM}
    \mathcal{N}^{S}= \Tr_E  \mathcal{U}_{S E}^n\circ\ldots \ldots\circ \mathcal{U}_{S E}^1\circ\mathcal{A}.
\end{equation}
The Choi state of $\mathcal{N}^{S}$ can be obtained by linking the process tensor $|\mathcal{T}_n)$ as follows:
\begin{equation}\label{MTUM}
    |\mathcal{T}^S)=N^{2n-2}((\prod_{i=1}^{n-1}  \Phi^{A^{i+1}S^i}) |\mathcal{T}_n). 
\end{equation}
The evolution $\mathcal{N}^{S}$ does not contain any measurements or control operations. If needed, one can apply CPTP operations $\mathcal{A}_i$  between the steps for manipulation purposes at intermediary time steps. This results in the following quantum channel:\begin{equation}\label{EQC}
    \mathcal{N}^{S}_{\textbf{A}_{n-1:1}}= \Tr_E  \mathcal{U}_{S E}^n\circ(\mathcal{A}_{n-1} \mathcal{U}_{S E}^{n-1})\circ\ldots \ldots\circ(\mathcal{A}_1 \mathcal{U}_{S E}^1)\circ\mathcal{A}.
\end{equation}
We graphically illustrate the many-body quantum channel and its derived channel in \cref{LINK}.

\begin{figure*}[htb]
    \centering
    \includegraphics[width=0.9\textwidth]{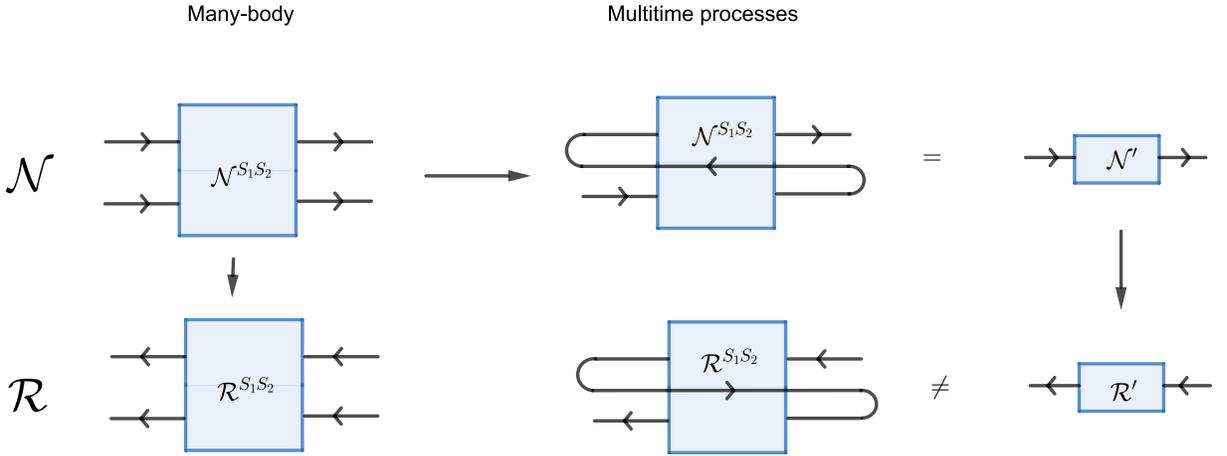}
    \caption{A many-body channel is linkable, which does not mean that its Petz recovery map is also linkable.
    The Petz recovery map of the derived channel cannot be obtained with the Petz recovery map of the corresponding many-body channel. 
    }\label{MBTS}
    \end{figure*}

One can derive channels by linking the indices of the many-body quantum channel:\begin{equation}\label{ECLK}
    \mathcal{N}' | \Phi^{A^{i}S^{i+1}})=N^2 (\Phi^{A^{i+1}S^i}|\mathcal{N}|\Phi^{A^{i}S^i}\otimes\Phi^{A^{i+1}S^{i+1}}).
\end{equation}
Therefore, it is natural to wonder whether the corresponding FTs can be obtained from the FTs for a many-body quantum channel. Unfortunately, this is not possible. A many-body quantum channel is time-ordered and linkable, and this does not necessarily mean that the corresponding Petz recovery is time-ordered and linkable. Therefore, the Petz recovery map of a derived channel cannot be obtained by linking the indices of the Petz recovery map of the many-body quantum channel. 
\begin{equation}
    ( \Phi^{A^{i}S^{i+1}}|\mathcal{R}'\neq N^2 (\Phi^{A^{i}S^i}\otimes\Phi^{A^{i+1}S^{i+1}}|\mathcal{R}|\Phi^{A^{i+1}S^i}).
\end{equation}
Consequently, one cannot obtain all those FTs in one step. These relations are depicted in \cref{MBTS}. 

As we cannot obtain all the FTs at one time, we analyze these channels separately. In this section, we first give the FTs for two ordinary channels. After that, we study the FTs for Markovian evolution, which allows for the complete measurement of intermediate states. However, these procedures are not applicable to non-Markovian evolution. We analyze the underlying causes of this fact. For non-Markovian evolution, we use a general measurement for the intermediate state and obtain the FTs that contain the effects of intermediate measurements.

\subsection{Two ordinary channels}\label{TOC}
Here, we present the FTs for many-body channels and channels that evolve without intermediate measurements. Both many-body channels $\mathcal{N}^{S^1\dots S^n}$ and $\mathcal{N}^S$ in \cref{EWIM} still follow single-step evolution. Therefore, the approaches and results shown in \cref{PFTWS} are applicable.

The many-body channels $\mathcal{N}^{S^1\dots S^n}$ map the initial tensor product states $\rho_I=\rho_I^{S_1}\otimes\dots \otimes  \rho_I^{S_n} =\sum_{u_1\dots u_n} \prod_{i=1}^n p_{u_i}^i   \Pi_{\psi_{u_i}}^{S_i}$ to their final states $\rho_F=\rho_F^{S_1\dots S_n} =\sum_{v'}  p'_{v'}   \Pi_{\phi_{v'}}^{S_1\dots S_n}$, which are not of tensor product form unless the evolution process is Markovian. The forward transition matrices can be defined as
\begin{equation}
    T_{{i_1j_1\dots i_nj_n }\to{k'l'}} 
    =N^n (\Pi_{i_1j_1}^{A_1}\otimes\dots\otimes\Pi_{i_nj_n}^{A_n}\otimes\Pi_{k'l'}^{S_1\dots S_n} | \mathcal{T}_n).
\end{equation}
The TPM quasiprobability distribution for the forward process can be defined as
\begin{align}
    P^{u_1\dots u_n, v'}_{i_1j_1\dots i_n j_n, k'l'}=T_{{i_1j_1\dots i_nj_n }\to{k'l'}} \times \notag \\
     (( \prod_{i=1}^n p_{u_i}^i  \Pi_{\psi_{u_i}}^{A_i})\otimes \Pi_{\phi'_{v'}}^{S_1\dots S_n}|\Pi_{i_1j_1}^{A_1}\otimes\dots\otimes\Pi_{i_nj_n}^{A_n}\otimes\Pi_{k'l'}^{S_1\dots S_n}).
\end{align}
The entropy production can be defined as
\begin{align}
    \sigma^{u_1\dots u_n \to v'}_{i_1j_1\dots i_nj_n\to k'l'}=-\log( {p'}_{v'})+\log(p^1_{u_1})+\dots +\log(p^n_{u_n}) \notag \\
    -(-\log(Z_{i_1j_1}\dots Z_{i_nj_n} Z'_{k'l'})).
\end{align}
The quantities for the backward process can be similarly defined. The relation in \cref{FTR} still holds for the quantities defined here, and the generalized second law becomes
\begin{align}
    \braket{\sigma}=S(\rho_I^{S_1\dots S_n}||\gamma_I^{S_1\dots S_n}) \notag \\
    -S(\mathcal{N}^{S_1\dots S_n}(\rho_I^{S_1\dots S_n})||\mathcal{N}^{S_1\dots S_n}(\gamma_I^{S_1\dots S_n}))\geq 0.
\end{align}

The channels that evolve without intermediate measurements give the Choi state in \cref{MTUM}. 
For simplicity, here, we consider only the two-step evolution $\mathcal{N}^S_{12}$, which maps the initial state $\rho_I^S$ to $\rho_F^{S}$. The corresponding generalized second law is\begin{equation}\label{NMTRS}
    \braket{\sigma}=S(\rho_I^S||\gamma^S)-S(\rho_F^{S}||\gamma^{S}_F).
\end{equation}
Suppose that $\mathcal{N}_2\circ\mathcal{N}_1$ is the Markov process that is closest to process $\mathcal{N}^S_{12}$ \cite{PRFPM18}. We can rewrite \cref{NMTRS} as
\begin{align}\label{HNMEFT}
    \braket{\sigma}=S(\rho_I^S||\gamma^S)- S(\mathcal{N}_1(\rho_I^S)||\mathcal{N}_1(\gamma^S))\notag \\
     +S(\rho_M^S||\gamma^S_M)-S(\mathcal{N}_2(\rho_M^S)||\mathcal{N}_2(\gamma^S_M))-\sigma_{NM},
\end{align}
where $\rho_M^S=\mathcal{N}_1(\rho_I^S)$. The nonnegative quantity
\begin{equation}\label{BLPM}
    \sigma_{NM}= S(\rho_F^{S}||\gamma^{S}_F)-S(\mathcal{N}_2(\rho_M^S)||\mathcal{N}_2(\gamma^S_M))
\end{equation}
is related to the increase in distinguishability, which implies the memory effects. The quantity (\ref{BLPM})  is similar to the quantity used in the Breuer-Laine-Piilo (BLP) measure \cite{BLP09}. \cref{HNMEFT} tells us that memory effects can reduce the system fluctuations.

\subsection{Markovian process}\label{MPTS}

The Petz recovery map of a Markovian process is time-ordered and linkable. The measurement of the intermediate state does not affect the future evolution. These properties make Markovian processes special. 

In a Markovian process, the system state evolves from the initial state $\rho_I=\sum_{u} p_{u}   \Pi_{\psi_{u}}^{S}$ to the intermediate state $\rho_M^i= \mathcal{N}_i \circ \dots \circ  \mathcal{N}_1( \rho_I)=\sum_{\mu_i} p_{\mu_i}   \Pi_{\mu_i}^{S}$ and finally to $\rho_F=\mathcal{N}_n \circ \dots \circ  \mathcal{N}_1( \rho_I)=\sum_{v'} p'_{v'}   \Pi_{\phi'_{v'}}^{S}$. We can completely measure the intermediate state. The linking operation $ |\Phi^{A^{i+1}S^i})$ in \cref{ECLK} is the Choi state of the identity mapping $ |\Phi^{A^{i+1}S^i})=I^{S} |\Phi^{A^{i+1}S^i})$. If we replace this operation with projection measurements $\mathcal{M}_\mu$, where $ \mathcal{M}_\mu (\rho)=\Pi_\mu (\rho) \Pi_\mu^\dagger$, we obtain the operation
\begin{equation}\label{LAMO}
    \mathcal{M}_\mu^{S^i}| \Phi^{A^{i+1}S^i})=\sum_\nu |\Pi^{A^{i+1}}_{\mu\nu}\otimes \Pi^{S^i}_{\mu\nu})/N.
\end{equation}
If such an operation is applied to the full intermediate state, we obtain a tensor similar to that in \cref{MTUM}: 
\begin{align}\label{FORM3}
    |\mathcal{T}_{\mu_1\dots \mu_{n-1}})=N^{n-1}\times \notag \\
    \sum_{\nu_1\dots \nu_{n-1}}((\prod_{i=1}^{n-1}  \Pi^{A^{i+1}}_{\mu_i\nu_i}\otimes \Pi^{S^i}_{\mu_i\nu_i})|\mathcal{N}^{S^1\dots S^n}| (\prod_{i=1}^n \Phi^{A^iS^i})). 
\end{align}
When we ignore the intermediate measurements, this tensor returns the unmeasured tensor: 
\begin{equation}
    \sum_{\mu_1\dots \mu_{n-1}}  |\mathcal{T}_{\mu_1\dots \mu_{n-1}})=|\mathcal{T}^S).
 \end{equation}

Now, we prove the FT. For simplicity, we only consider the two-step evolution process. The system state evolves from $\rho_I=\sum_u p_u   \Pi_{\psi_u}$ to $\rho_M=\sum_{\mu} p'_{\mu}   \Pi_{\mu}$ and finally to $\rho_F=\sum_{w''} p''_{w''}   \Pi_{{\xi''}_{w''}}$. Since the many-body channel of Markovian evolution gives tensor product states, we can define
the forward transition matrices as
 \begin{widetext}
 \begin{equation}
   T_{ij,kl\to k'l',m'n'}=N^2 ( \Pi_{ij}^A\otimes\Pi_{k'l'}^S\otimes\Pi_{kl}^{A'}\otimes\Pi_{m'n'}^{S'}|\mathcal{N}^{S'}_2\circ\mathcal{N}^S_1|\Phi^{AS}\otimes \Phi^{A'S'}).
\end{equation}
As in the procedure of \cref{jgn}, the forward transition matrices can be expressed with a Petz recovery map:
    \begin{equation}
        T_{ij,kl\to k'l',m'n'}^*=N^2( \Phi^{AS}\otimes \Phi^{A'S'}|   \mathcal{R}_{\gamma,\gamma'}| (\mathcal{J}_{\gamma,\gamma',\mathcal{N}}^{ASA'S'})^{1/2}\Pi_{ij}^A\otimes\Pi_{k'l'}^S\otimes\Pi_{kl}^{A'}\otimes\Pi_{m'n'}^{S'}),
    \end{equation}
where the Petz recovery map $\mathcal{R}_{\gamma,\gamma'}= (\mathcal{J}_\gamma^S)^{1/2}\circ(\mathcal{J}_{\gamma'}^{S'})^{1/2}\circ {\mathcal{N}^S_1}^\dagger \circ {\mathcal{N}^{S'}_2}^\dagger  \circ  (\mathcal{J}_{\mathcal{N}(\gamma^S\otimes \gamma'^{S'})}^{SS'})^{-1/2}$ is a CPTP map and $ \mathcal{J}_{\gamma,\gamma',\mathcal{N}}^{ASA'S'}=(\mathcal{J}_\gamma^A)^{-1}\otimes (\mathcal{J}_{\gamma'}^{A'})^{-1}\otimes \mathcal{J}_{\mathcal{N}(\gamma^S\otimes \gamma'^{S'})}^{SS'}$. The final reference state here is the tensor product state $\mathcal{N}(\gamma^S\otimes \gamma'^{S'})=\mathcal{N}^S_1(\gamma^S)\otimes \mathcal{N}^{S'}_2( \gamma'^{S'})$, which makes the Petz recovery map divisible: $\mathcal{R}_{\gamma,\gamma'}= {\mathcal{R}^S_1} \circ{\mathcal{R}^{S'}_2}$. The factor of the rescaled operators becomes
    \begin{align}
        Z^{{\gamma}^{-1}}_{ij}Z^{{\gamma'}^{-1}}_{kl}Z^{\mathcal{N}^S_1(\gamma^S)}_{k'l'} Z^{\mathcal{N}^{S'}_2(\gamma^{S'})}_{m'n'}:=\norm{ (\mathcal{J}_{\gamma,\gamma',\mathcal{N}}^{ASA'S'})^{1/2}\Pi_{ij}^A\otimes\Pi_{k'l'}^S\otimes\Pi_{kl}^{A'}\otimes\Pi_{m'n'}^{S'}}_2\notag \\
        =\norm{\mathcal{J}_{\gamma}^{-1/2}\Pi_{ij}}_2\times\norm{\mathcal{J}_{\gamma'}^{-1/2}\Pi_{kl}}_2\times \norm{\mathcal{J}_{\mathcal{N}^S_1(\gamma^S)}^{1/2}\Pi_{k'l'}^{S}}_2 \times \norm{\mathcal{J}_{\mathcal{N}^{S'}_2(\gamma^{S'})}^{1/2} \Pi_{m'n'}^{S'}}_2.
     \end{align}
 The backward transition matrices can be defined as
     \begin{equation}
        \tilde{T}_{ij,kl\leftarrow k'l',m'n'}=N^2( \Phi^{AS}\otimes \Phi^{A'S'}| \mathcal{R}_{\gamma,\gamma'}|  {\Pi'}_{ij}^A\otimes{\Pi'}_{k'l'}^S\otimes{\Pi'}_{kl}^{A'}\otimes{\Pi'}_{m'n'}^{S'}).
     \end{equation}
     where $|  {\Pi'}_{ij}^A\otimes{\Pi'}_{k'l'}^S\otimes{\Pi'}_{kl}^{A'}\otimes{\Pi'}_{m'n'}^{S'})=| (\mathcal{J}_{\gamma,\gamma',\mathcal{N}}^{ASA'S'})^{1/2}\Pi_{ij}^A\otimes\Pi_{k'l'}^S\otimes\Pi_{kl}^{A'}\otimes\Pi_{m'n'}^{S'})/(Z^{{\gamma}^{-1}}_{ij}Z^{{\gamma'}^{-1}}_{kl}Z^{\mathcal{N}^S_1(\gamma^S)}_{k'l'} Z^{\mathcal{N}^{S'}_2(\gamma^{S'})}_{m'n'})$. The relation between the forward transition matrices and the backward transition matrices is 
     \begin{equation}
        T_{ij,kl\to k'l',m'n'}=  \tilde{T}_{ij,kl\leftarrow k'l',m'n'}^*\times (Z^{{\gamma}^{-1}}_{ij}Z^{{\gamma'}^{-1}}_{kl}Z^{\mathcal{N}^S_1(\gamma^S)}_{k'l'} Z^{\mathcal{N}^{S'}_2(\gamma^{S'})}_{m'n'}).
     \end{equation}
     Since we need to measure the intermediate state, the TPM should be turned into a three-point measurement. According to \cref{FORM3}, we define the quasiprobability distribution of the three-point measurement for the forward process as \begin{align} \label{TMQDF}
         P^{u, \mu,w''}_{ij, k'l',kl,m'n'}=p_u \sum_\nu  N^2(   \Pi_{\psi_u}^A\otimes\Pi_{\mu\nu}^S\otimes  \Pi_{\mu\nu}^{A'}\otimes  \Pi_{{\xi''}_{w''}}^{S'}|\Pi_{ij}^A\otimes\Pi_{k'l'}^S\otimes\Pi_{kl}^{A'}\otimes\Pi_{m'n'}^{S'}) \notag \\
        \times (\Pi_{ij}^A\otimes\Pi_{k'l'}^S\otimes\Pi_{kl}^{A'}\otimes\Pi_{m'n'}^{S'}| {\mathcal{N}^{S'}_2} \circ{\mathcal{N}^S_1}|\Phi^{AS}\otimes \Phi^{A'S'}).
     \end{align}
We are free to choose the basis of $\Pi_{\mu\nu}^S$ and $\Pi_{\mu\nu}^{A'}$. Here, we set it equal to the basis of $ \{ \Pi_{{\phi'}_{v'}}\}$. It is easy to prove that $P^{u, \mu,w''}_{ij, k'l',kl,m'n'}$ satisfies the marginality condition:
     \begin{align} \label{MTMQDF}
        \sum_{\mu,w'',ij, k'l',kl,m'n'} P^{u, \mu,w''}_{ij, k'l',kl,m'n'}= p_u N^3(   \Pi_{\psi_u}^A\otimes \Phi^{SA'}\otimes  I^{S'}|{\mathcal{N}^{S'}_2} \circ{\mathcal{N}^S_1}|\Phi^{AS}\otimes \Phi^{A'S'})
        =p_u(    I^S  |{\mathcal{N}^{S}_2} \circ{\mathcal{N}^S_1}|  \Pi_{\psi_u}^S)=p_u , \notag\\
         \sum_{u,w'',ij, k'l',kl,m'n'}  P^{u, \mu,w''}_{ij, k'l',kl,m'n'}=\sum_{\nu} N^2(     \rho_I^A\otimes \Pi_{\mu\nu}^S\otimes  \Pi_{\mu\nu}^{A'}\otimes I^{S'}|{\mathcal{N}^{S'}_2} \circ{\mathcal{N}^S_1}|\Phi^{AS}\otimes \Phi^{A'S'})\notag \\
         =\sum_{\nu} (  \Pi_{{\xi''}_{w''}}^{S} |{\mathcal{N}^{S}_2} | \Pi_{\mu\nu}^S)(  \Pi_{\mu\nu}^S|{\mathcal{N}^S_1}| \rho_I^S) ={p'}_{\mu} ,  \notag\\
        \sum_{u,\mu,ij, k'l',kl,m'n'}  P^{u, \mu,w''}_{ij, k'l',kl,m'n'}=\sum_{\mu\nu} N^2(     \rho_I^A\otimes \Pi_{\mu\nu}^S\otimes  \Pi_{\mu\nu}^{A'}\otimes \Pi_{{\xi''}_{w''}}^{S'}|{\mathcal{N}^{S'}_2} \circ{\mathcal{N}^S_1}|\Phi^{AS}\otimes \Phi^{A'S'})\notag \\
        =\sum_{\mu\nu} (  \Pi_{{\xi''}_{w''}}^{S} |{\mathcal{N}^{S}_2} | \Pi_{\mu\nu}^S)(  \Pi_{\mu\nu}^S|{\mathcal{N}^S_1}| \rho_I^S) =  (  I^{S} |{\mathcal{N}^{S}_2} \circ{\mathcal{N}^S_1}| \rho_I^S) ={p''}_{w''} , \notag\\
         \sum_{u,\mu,w'',kl, k'l',m'n'}   P^{u, \mu,w''}_{ij, k'l',kl,m'n'}=N^3(     \rho_I^A\otimes \Phi^{SA'}\otimes  I^{S'}|\Pi_{ij}^{A})(\Pi_{ij}^{A}|{\mathcal{N}^{S'}_2} \circ{\mathcal{N}^S_1}|\Phi^{AS}\otimes \Phi^{A'S'})\notag \\
        = (  I^{S} |{\mathcal{N}^{S'}_2} \circ{\mathcal{N}^S_1}|\Pi_{ij}^{S})(\Pi_{ij}^{S}| \rho_I^S) = \delta_{ij}(\Pi_{ij}^{S}| \rho_I^S),  \notag\\
        \sum_{u,\mu,w'',ij, k'l',m'n'}   P^{u, \mu,w''}_{ij, k'l',kl,m'n'}=N^3(     \rho_I^A\otimes \Phi^{SA'}\otimes  I^{S'}|\Pi_{kl}^{A'})(\Pi_{kl}^{A'}|{\mathcal{N}^{S'}_2} \circ{\mathcal{N}^S_1}|\Phi^{AS}\otimes \Phi^{A'S'})\notag \\
       = (  I^{S'} |{\mathcal{N}^{S'}_2} | \Pi_{kl}^{S'})(  \Pi_{kl}^{S}|{\mathcal{N}^S_1}| \rho_I^S) = \delta_{kl}(  \Pi_{kl}^{S}| \rho_M^S)  , \notag\\
        \sum_{u,\mu,w'',ij, kl,m'n'}   P^{u, \mu,w''}_{ij, k'l',kl,m'n'}=N^3(     \rho_I^A\otimes \Phi^{SA'}\otimes  I^{S'}|\Pi_{k'l'}^S)(\Pi_{k'l'}^S|{\mathcal{N}^{S'}_2} \circ{\mathcal{N}^S_1}|\Phi^{AS}\otimes \Phi^{A'S'})\notag \\
       = (\Pi_{k'l'}^S\otimes I^{S'}|{\mathcal{N}^{S'}_2} \circ{\mathcal{N}^S_1}| \rho_I^S\otimes \Pi_{k'l'}^{S'} ) =( I^{S'}|{\mathcal{N}^{S'}_2} |\Pi_{k'l'}^{S'}) (  \Pi_{k'l'}^{S}| \rho_M^{S})=   \delta_{k'l'}(  \Pi_{k'l'}^{S}| \rho_M^S) , \notag\\
        \sum_{u,\mu,w'',ij,k'l', kl}   P^{u, \mu,w''}_{ij, k'l',kl,m'n'}=N^3(     \rho_I^A\otimes \Phi^{SA'}\otimes  I^{S'}|\Pi_{m'n'}^{S'})(\Pi_{m'n'}^{S'}|{\mathcal{N}^{S'}_2} \circ{\mathcal{N}^S_1}|\Phi^{AS}\otimes \Phi^{A'S'})\notag \\
       =\delta_{m'n'}(\Pi_{m'n'}^{S}|{\mathcal{N}^{S}_2} \circ{\mathcal{N}^S_1}| \rho_I^S) =\delta_{m'n'}(\Pi_{m'n'}^{S}| \rho_F^S) .
     \end{align}
     The entropy production can be defined as
     \begin{equation}\label{SIGMP}
         \sigma^{u\to\mu\to w''}_{ijkl\to k'l'm'n'}=(\delta s^{u\to\mu}-\delta q_{ij\to k'l'})+(\delta s^{\mu\to w''}-\delta q_{kl\to m'n'}),
     \end{equation}
    where $\delta s^{u\to\mu}=-\log( {p'}_{\mu})+\log({p}_{u})$, $\delta s^{\mu\to w''}=-\log( {p''}_{w''})+\log({p'}_{\mu})$, $\delta q_{ij\to k'l'}=-\log(Z^{{\gamma}^{-1}}_{ij}Z^{\mathcal{N}^S_1(\gamma^S)}_{k'l'}) $ and $\delta q_{kl\to m'n'}=-\log(Z^{{\gamma'}^{-1}}_{kl}Z^{\mathcal{N}^{S'}_2(\gamma^{S'})}_{m'n'})$. The entropy production distribution $\sigma$ is
     \begin{equation}\label{DEPF}
         P_\to(\sigma)=\sum_{u,i,j}\sum_{\mu,k,l}\sum_{w'',k',l',m',n'}  P^{u, \mu,w''}_{ij, k'l',kl,m'n'}\delta(\sigma- \sigma^{u\to\mu\to w''}_{ijkl\to k'l'm'n'})
     \end{equation}
   The quasiprobability distribution of the three-point measurement for the backward process can be defined as
     \begin{align}\label{TMQDB}
         {P'}^{u, \mu,w''}_{ij, k'l',kl,m'n'}=p''_{w''} \sum_\nu  N^2 (\Phi^{AS}\otimes \Phi^{A'S'}| {\mathcal{R}^S_1} \circ{\mathcal{R}^{S'}_2}|{\Pi'}_{ij}^A\otimes{\Pi'}_{k'l'}^S\otimes{\Pi'}_{kl}^{A'}\otimes{\Pi'}_{m'n'}^{S'})\notag \\
        \times( \Pi_{ij}^A\otimes\Pi_{k'l'}^S\otimes\Pi_{kl}^{A'}\otimes\Pi_{m'n'}^{S'} | \Pi_{\psi_u}^A\otimes\Pi_{\mu\nu}^S\otimes  \Pi_{\mu\nu}^{A'}\otimes  \Pi_{{\xi''}_{w''}}^{S'}) .
     \end{align}
     $ {P'}^{u, \mu,w''}_{ij, k'l',kl,m'n'}$ also satisfies the marginality condition:
     \begin{align}
         \sum_{u,\mu,ij,k'l', kl,m'n'}   {P'}^{u, \mu,w''}_{ij, k'l',kl,m'n'}= p''_{w''}  N^2 (I^{S}\otimes \Phi^{A'S'}|{\mathcal{R}^S_1} \circ{\mathcal{R}^{S'}_2} |\Phi^{SA'}\otimes  \Pi_{{\xi''}_{w''}}^{S'})
         = p''_{w''}   (I^{S}|{\mathcal{R}^S_1} \circ{\mathcal{R}^{S}_2} |  \Pi_{{\xi''}_{w''}}^{S})=p''_{w''} , \notag \\ 
         \sum_{u,w'',ij,k'l', kl,m'n'}   {P'}^{u, \mu,w''}_{ij, k'l',kl,m'n'}= \sum_\nu N^2 (I^{S}\otimes \Phi^{A'S'}|{\mathcal{R}^S_1} \circ{\mathcal{R}^{S'}_2} |\Pi_{\mu\nu}^S\otimes  \Pi_{\mu\nu}^{A'}\otimes  \rho_F^{S'})
         = ( \Pi_{\mu}^{S'}|{\mathcal{R}^{S'}_2} |    \rho_F^{S'}), \notag \\ 
         \sum_{\mu,w'',ij,k'l', kl,m'n'}   {P'}^{u, \mu,w''}_{ij, k'l',kl,m'n'}= \sum_{\mu\nu} N^2 ( \Pi_{\psi_u}^S\otimes \Phi^{A'S'}|{\mathcal{R}^S_1} \circ{\mathcal{R}^{S'}_2} |\Pi_{\mu\nu}^S\otimes  \Pi_{\mu\nu}^{A'}\otimes  \rho_F^{S'})
         = ( \Pi_{\psi_u}^S|{\mathcal{R}^{S}_1} |   \mathcal{R}^{S}_2( \rho_F^{S})).
     \end{align}
     The relation in \cref{FTR} still holds with the quantities defined here. Combining \cref{MTMQDF,SIGMP,DEPF}, we obtain
     \begin{equation}\label{MTRS}
         \braket{\sigma}=S(\rho_I||\gamma)-S(\rho_M||{\mathcal{N}^S_1}(\gamma))+S(\rho_M||\gamma')-S(\rho_F||{\mathcal{N}^S_2}(\gamma'))
     \end{equation}
     Different from the previous result (\ref{NMTRS}), the average entropy production in \cref{MTRS} contains the intermediate state $\rho_M$. Therefore, a change in the intermediate state will affect the entropy production defined here. The FTs described here are extensions of the previous procedure. If we choose $\gamma'={\mathcal{N}^S_1}(\gamma)$, then the average entropy production returns to the previous result:
     \begin{equation}\label{RTUMF}
          \braket{\sigma}=S(\rho_I||\gamma)-S(\rho_F||{\mathcal{N}^S_2}(\gamma'))=S(\rho_I||\gamma)-S({\mathcal{N}^S_2}\circ{\mathcal{N}^S_1}(\rho_I)||{\mathcal{N}^S_2}\circ{\mathcal{N}^S_1}(\gamma)).
     \end{equation}
     The freely chosen intermediate reference state $\gamma'$ can bring some convenience. For example, the reference state is often selected from the global fixed points of the quantum channel. In multitime processes, the problem is that one cannot ensure that the evolved reference state $\mathcal{N}_t(\gamma)$ is always the global fixed point. In contrast, the extra reference state $\gamma'$ can always be selected from the global fixed points of $\mathcal{N}^S_2$. Therefore, 
the method proposed in this section is more suitable for multitime processes.

    Even for the cases in which $\gamma'={\mathcal{N}^S_1}(\gamma)$, \cref{MTRS} has deeper meaning. The quantity in (\ref{SIGMP}) is the composite of two parts, both of which satisfy the fluctuation relation. $\sigma_1:=\delta s^{u\to \mu}-\delta q_{ij\to k'l'}$ is the entropy production of quantum channel $\mathcal{N}_1$ according to the time-ordered property. Hence, its fluctuation relation is obvious from \cref{PFTWS}. For $\sigma_2:=\delta s^{\mu\to w''}-\delta q_{kl\to m'n'}$, the distribution can be derived from \cref{DEPF}:
     \begin{equation}
         P_\to(\sigma_2)=\sum_{\mu, w'',k',l',m',n'} (\sum_{u,i,j,k,l} P^{u, \mu,w''}_{ij, k'l',kl,m'n'})\delta(\sigma- \sigma_2)=\sum_{\mu, w'',k',l',m',n'}  P^{\mu, w''}_{kl,m'n'}\delta(\sigma- \sigma_2)
     \end{equation}
     Combining this with \cref{TMQDF}, we obtain
     \begin{align}
         P^{\mu, w''}_{kl,m'n'}=\sum_\nu N^2(   \rho_I^A\otimes\Pi_{\mu\nu}^S\otimes  \Pi_{\mu\nu}^{A'}\otimes  \Pi_{{\xi''}_{w''}}^{S'}|\Pi_{kl}^{A'}\otimes\Pi_{m'n'}^{S'}) 
          (\Pi_{kl}^{A'}\otimes\Pi_{m'n'}^{S'}| {\mathcal{N}^{S'}_2} \circ{\mathcal{N}^S_1}|\Phi^{AS}\otimes \Phi^{A'S'})\notag \\
        = \sum_\nu N(   \Pi_{\mu\nu}^S\otimes  \Pi_{\mu\nu}^{A'}\otimes  \Pi_{{\xi''}_{w''}}^{S'}|\Pi_{kl}^{A'}\otimes\Pi_{m'n'}^{S'}) 
        (\Pi_{kl}^{A'}\otimes\Pi_{m'n'}^{S'}| {\mathcal{N}^{S'}_2} |\rho_M^{S}\otimes \Phi^{A'S'}) \notag\\
        =p'_{\mu}N (  \Pi_{\mu}^{A'}\otimes  \Pi_{{\xi''}_{w''}}^{S'}|\Pi_{kl}^{A'}\otimes\Pi_{m'n'}^{S'}) 
        (\Pi_{kl}^{A'}\otimes\Pi_{m'n'}^{S'}| {\mathcal{N}^{S'}_2} |\Phi^{A'S'}) .
     \end{align}
     Comparing this with \cref{TPMQDF}, we know that $\sigma_2$ also satisfies the fluctuation relation $ P_\to(\sigma_2)/{P_\leftarrow(-\sigma_2)}=e^{\sigma_2}$. The generalized second law gives $ \braket{\sigma_2}=S(\rho_M||\gamma')-S(\rho_F||{\mathcal{N}^S_2}(\gamma'))$.

\subsection{The conflict between intermediate measurements and non-Markovian processes}
 In the above section, we extend the FTs for non-Markovian processes. We realize intermediate measurements with the operation in (\ref{LAMO}) and prove that the FTs contain the intermediate state of the system. It is natural to ask whether this procedure is applicable to non-Markovian processes. The answer is no, and we discuss this from different perspectives.

We first discuss the conflict through concrete examples. We still consider the two-step evolution process here. Since the considered evolution process is non-Markovian, the final state of the two-body channel is not of tensor product form. The forward transition matrices should be  
\begin{equation}
    T_{ij,kl\to k'l',m'n'}=N^2  (\Pi_{ij}^A\otimes\Pi_{kl}^{A'}\otimes\Pi_{m'n'}^{SS'}|{\mathcal{N}^{SS'}}|\Phi^{AS}\otimes \Phi^{A'S'}).
\end{equation}
If we still define the quasiprobability distribution of the three-point measurement according to \cref{FORM3}, we obtain 
\begin{align}
    P^{u, \mu,w''}_{ij, k'l',kl,m'n'}=p_u \sum_\nu  N^2(   \Pi_{\psi_u}^A\otimes\Pi_{\mu\nu}^S\otimes  \Pi_{\mu\nu}^{A'}\otimes  \Pi_{{\xi''}_{w''}}^{S'}|\Pi_{ij}^A\otimes\Pi_{kl}^{A'}\otimes\Pi_{m'n'}^{SS'}) \notag \\
   \times (\Pi_{ij}^A\otimes\Pi_{kl}^{A'}\otimes\Pi_{m'n'}^{SS'}|{\mathcal{N}^{SS'}}|\Phi^{AS}\otimes \Phi^{A'S'}).
\end{align}
The main problem comes from the fact that the Petz recovery map
\begin{equation}
    {\mathcal{R}^{SS'}_{\gamma\otimes\gamma'}}=\mathcal{J}_{\gamma^S\otimes{\gamma'}^{S'}}^{1/2}\circ{\mathcal{N}^{SS'}}^\dagger\circ\mathcal{J}_{\mathcal{N}^{SS'}(\gamma^S\otimes{\gamma'}^{S'})}^{-1/2}
\end{equation}
is neither time-ordered nor linkable. These issues cause the quasiprobability distribution for the backward process
\begin{align}\label{NMTMQDB}
    {P'}^{u, \mu,w''}_{ij, k'l',kl,m'n'}=p''_{w''} \sum_\nu  N^2  (\Phi^{AS}\otimes \Phi^{A'S'}|{\mathcal{R}^{SS'}_{\gamma\otimes\gamma'}}|{\Pi'}_{ij}^A\otimes{\Pi'}_{k'l'}^S\otimes{\Pi'}_{kl}^{A'}\otimes{\Pi'}_{m'n'}^{S'})\notag \\
   \times(  {\Pi'}_{ij}^A\otimes{\Pi'}_{k'l'}^S\otimes{\Pi'}_{kl}^{A'}\otimes{\Pi'}_{m'n'}^{S'} |\Pi_{\psi_u}^A\otimes\Pi_{\mu\nu}^S\otimes  \Pi_{\mu\nu}^{A'}\otimes  \Pi_{{\xi''}_{w''}}^{S'})
\end{align}
to not satisfy the marginality condition:
\begin{align}
    \sum_{u,w'',ij, k'l',kl,m'n'}  {P'}^{u, \mu,w''}_{ij, k'l',kl,m'n'}=\sum_{\nu} N^2(  \Phi^{AS}\otimes \Phi^{A'S'}   | {\mathcal{R}^{SS'}_{\gamma\otimes\gamma'}}|I^A\otimes \Pi_{\mu\nu}^S\otimes  \Pi_{\mu\nu}^{A'}\otimes \rho_F^{S'})\notag \\
    =\sum_{\nu} (  I^{S}\otimes \Pi_{\mu\nu}^{S'} | {\mathcal{R}^{SS'}_{\gamma\otimes\gamma'}}| \Pi_{\mu\nu}^S \otimes \rho_F^{S'}) \neq p'_\mu, \notag \\ 
    \sum_{\mu,w'',ij, k'l',kl,m'n'}  {P'}^{u, \mu,w''}_{ij, k'l',kl,m'n'}=\sum_{\mu\nu} N^2(  \Phi^{AS}\otimes \Phi^{A'S'}   | {\mathcal{R}^{SS'}_{\gamma\otimes\gamma'}}|\Pi_{\psi_u}^A\otimes \Pi_{\mu\nu}^S\otimes  \Pi_{\mu\nu}^{A'}\otimes \rho_F^{S'})\notag \\
    =\sum_{\mu\nu} ( \Pi_{\psi_u}^{S}\otimes \Pi_{\mu\nu}^{S'} | {\mathcal{R}^{SS'}_{\gamma\otimes\gamma'}}| \Pi_{\mu\nu}^S \otimes \rho_F^{S'}) \neq  p'_u  .
 \end{align}
Hence, the quasiprobability distribution in \cref{NMTMQDB} is ill-defined. The procedure of \cref{MPTS} is not suitable for non-Markovian processes.
 
\end{widetext}
On the other hand, TPMs need to know the complete information about the initial and final states, so they are complete measurements \cite{DPS04}. �Complete measurements� mean that we can copy or broadcast these states, which requires that these states can be maximally entangled with an auxiliary state \cite{L06}. However, the final state $\rho_f^{S_1\dots S_n} $ of a non-Markovian process shows that correlations are present between the intermediate state and the other states. According to the exclusivity of entanglement, the complete measurements of the intermediate state conflict with a non-Markovian process.

These conclusions are consistent with the analysis in \cref{INTRO}. In a non-Markovian process, the measurements taken over the intermediate state influence the later evolution process and conflict with the FTs. To obtain the FTs for non-Markovian processes, one approach is to avoid measuring the intermediate states, as in \cref{TOC}. Another approach is to include a measurement component in the evolution process and use derived channels to deduce the FTs. 
In the following section, we show how to construct a suitable channel and prove the corresponding FTs.

\subsection{The FTs for non-Markovian processes}\label{NMCQPT}
As shown in \cref{EQC}, we can derive a quantum channel by inserting operations between the process steps. A simple and direct approach is to insert a projective measurement $\mathcal{A}=\sum_k \Pi_k(\cdot) \Pi_k$. However, such a measurement itself causes entropy production, which makes it difficult to separate the contributions of various components. In addition, we cannot obtain information about the intermediate states because the measured results are sent to the next step and not retained.

Here, we use the following unitary evolution
\begin{align}\label{UEO}
    \mathcal{A}^i=\mathcal{A}^i_{S S'_i} (\cdot\otimes \ket{0}_{S'_i}\bra{0})\notag \\
    =\mathcal{U}_{ S S'_i}[(\cdot)\otimes \ket{0}_{S'_i}\bra{0}]=\sum_{kl}  {\Pi_{k}^{S}} (\cdot) {\Pi_{l}^{S}}^\dagger  \otimes  \Pi_{kl}^{S'_i}
\end{align}
to realize general quantum measurements \cite{F95} for intermediate states. The operation $\mathcal{A}^i$ is a unitary transformation for $SS'_i$. This operation does not lead to entropy production. The ancillary $S'_i$ also records the probability distributions of the intermediate state with respect to the basis $ \{\Pi_{k}\}$, i.e., $\Tr_S  \mathcal{A}_{S\to S S'}(\rho_M^S)=\sum_{k}  \Pi_{k} \rho_M \Pi_{k}^\dagger$. The measurement here cannot obtain the complete information about the intermediate states. The correlations between the system and the environment or the process itself produce natural limitations regarding the available information. That is why the measurement procedure utilized here does not conflict with a non-Markovian process.

Here, we still consider the two-step evolution process. The derived channel 
\begin{equation}\label{ECOTSE}
    \mathcal{N}_{S S'}=\Tr_E \mathcal{U}_{SE}^{2} \circ\mathcal{A}_{ SS'} \circ\mathcal{U}_{SE}^{1}\circ\mathcal{A}
\end{equation}
 is a CPTP map for $SS'$, which can also be expressed as $ \mathcal{N}_{ S S'}=\sum_i M_i^{SS'}(\cdot ){M_i^{SS'}}^\dagger  $.
It maps the initial state $\rho_I=(\sum_u p_u   \Pi_{\psi_u}^{S})\otimes \Pi_0^{S'}$ to $\rho_F=\sum_{v'} {p'}_{v'}\Pi_{\phi'_{v'}}^{S S'}$. The forward transition matrices can be defined as
\begin{equation}
    T_{{ij}\to{k'l'}}=N  ( \Pi_{ij}^{A}\otimes\Pi_{k'l'}^{SS'}| \mathcal{N}_{ S S'} |\Phi^{AS}\otimes \Pi_0^{S'}).
\end{equation}
As in the procedure of \cref{jgn}, we have
\begin{equation}
    T_{{ij}\to{k'l'}}^*=N(\Phi^{AS}\otimes \Pi_0^{S'}|  \mathcal{R}_{SS'}^\gamma|  (\mathcal{J}_{\gamma,\mathcal{N}}^{ASS'})^{1/2} \Pi_{ij}^{A}\otimes\Pi_{k'l'}^{SS'}),
\end{equation}
where the Petz recovery map
\begin{equation}
    \mathcal{R}_{SS'}^\gamma=(\mathcal{J}_{\Pi_0}^{S'})^{1/2}\circ (\mathcal{J}_\gamma^{S})^{1/2}\circ\mathcal{N}_{SS'}^\dagger\circ (\mathcal{J}_{\mathcal{N}_{SS'}(\gamma^{SS'})}^{SS'})^{-1/2}.
\end{equation}
The rescaling map $ \mathcal{J}_{\gamma,\mathcal{N}}^{ASS'}=(\mathcal{J}_\gamma^{A})^{-1}\otimes \mathcal{J}_{\gamma_F}^{SS'}$, where $\gamma_F=\mathcal{N}_{SS'}(\gamma^{SS'})$ is the final reference state and the reference state $\gamma^{SS'}=\gamma^S\otimes  \Pi_0^{S'}$. The  trace preserving  property of this Petz recovery map is obvious from
\begin{align}
    (I^{SS'}|(\mathcal{J}_{\Pi_0}^{S'})^{1/2}\circ (\mathcal{J}_\gamma^{S})^{1/2}\circ\mathcal{N}_{SS'}^\dagger\circ (\mathcal{J}_{\gamma_F}^{SS'})^{-1/2} |O^{SS'}) \notag\\
    =(O^{SS'}| (\mathcal{J}_{\gamma_F}^{SS'})^{-1/2}\mathcal{N}_{SS'}|\gamma\otimes\Pi_0^{S'} )^*= \Tr(O).
\end{align}
The rescaling map $(\mathcal{J}_{\Pi_0}^{S'})^{1/2}$ allows the final states of $\mathcal{R}_{SS'}^\gamma$ to maintain the form ${\rho'}_F^S\otimes \Pi_0^{S'}$.
The factor of the rescaled operators becomes 
\begin{align}\label{FOTPcps}
   Z^{{\gamma}^{-1}}_{ij}  Z^{\gamma_F}_{k'l'}:=\norm{ ( \mathcal{J}_{\gamma,\mathcal{N}}^{ASS'})^{1/2} \Pi_{ij}^{A}\otimes\Pi_{k'l'}^{SS'}}_2\notag\\
   =\norm{\mathcal{J}_{\gamma}^{-1/2}\Pi_{ij}}_2\times \norm{\mathcal{J}_{\gamma_F}^{1/2} \Pi_{k'l'}^{SS'}}_2.
\end{align}
The backward transition matrices can be defined as
     \begin{equation}
        \tilde{T}_{{ij}\leftarrow{k'l'}}=N(\Phi^{AS}\otimes \Pi_0^{S'}|  \mathcal{R}_{SS'}^\gamma| {\Pi'}_{ij}^{A}\otimes{\Pi'}_{k'l'}^{SS'}),
     \end{equation}
where the reference-rescaled operators
\begin{equation}
    | {\Pi'}_{ij}^{A}\otimes{\Pi'}_{k'l'}^{SS'})=|  (\mathcal{J}_{\gamma,\mathcal{N}}^{ASS'})^{1/2} \Pi_{ij}^{A}\otimes\Pi_{k'l'}^{SS'})/(Z^{{\gamma}^{-1}}_{ij}  Z^{\gamma_F}_{k'l'}).
\end{equation}
The relation between the forward transition matrices and the backward transition matrices is 
\begin{equation}
    T_{{ij}\to{k'l'}}=\tilde{T}_{{ij}\leftarrow{k'l'}}^*\times (Z^{{\gamma}^{-1}}_{ij} Z^{\gamma_F}_{k'l'}).
\end{equation}
Since a general measurement of the intermediate state is realized with ancilla measurements, the TPM of the system should be turned into a TPM for the system ancilla. The quasiprobability distribution of the TPM for the forward process can be defined as
\begin{align}\label{PDTcps}
    P^{u, v'}_{ij, k'l'}=p_u  N(   \Pi_{\psi_u}^A\otimes \Pi_{\phi'_{v'}}^{SS'}|\Pi_{ij}^A\otimes\Pi_{k'l'}^{SS'}) \notag \\
   \times  ( \Pi_{ij}^{A}\otimes\Pi_{k'l'}^{SS'}| \mathcal{N}_{ S S'} |\Phi^{AS}\otimes \Pi_0^{S'}).
\end{align}
It is easy to prove that $ P^{u, v'}_{ij, k'l'}$ satisfies the marginality condition:
\begin{align}\label{COEPTcps}
    \sum_{v',ij,k'l'}  P^{u, v'}_{ij, k'l'}=p_u N(    \Pi_{\psi_u}^A\otimes  I^{SS'}|\mathcal{N}_{ S S'}|\Phi^{AS}\otimes \Pi_0^{S'}) \notag \\
    =p_u  (   I^{SS'}  |\mathcal{N}_{ S S'}| \Pi_{\psi_u}^S\otimes \Pi_0^{S'}) ={p}_u , \notag\\
   \sum_{u,ij, k'l'}  P^{u, v'}_{ij, k'l'}=N (     \rho_I^A\otimes\Pi_{\phi'_{v'}}^{SS'}|\mathcal{N}_{ S S'}|\Phi^{AS}\otimes \Pi_0^{S'}) \notag \\
   =(  \Pi_{\phi'_{v'}}^{SS'}|\rho_F^{SS'})={p'}_{v'} , \notag \\
    \sum_{u,v', k'l'}   P^{u, v'}_{ij, k'l'}=N(   \rho_I^A\otimes   I^{SS'}|\Pi_{ij}^{A})\notag \\
    \times(\Pi_{ij}^{A}|\mathcal{N}_{ S S'}|\Phi^{AS}\otimes \Pi_0^{S'} )= \delta_{ij}(\Pi_{ij}^{S}| \rho_I^S)  ,\notag \\
    \sum_{u,v', ij}   P^{u, v'}_{ij, k'l'}=N(   \rho_I^A\otimes   I^{SS'}|\Pi_{k'l'}^{SS'})\notag\\
   \times (\Pi_{k'l'}^{SS'}|\mathcal{N}_{ S S'}|\Phi^{AS}\otimes \Pi_0^{S'})=\delta_{k'l'}(\Pi_{m'n'}^{SS'}| \rho_F^{SS'}).
\end{align}
The entropy production can be defined as
\begin{equation}\label{SIGOTPcps}
    \sigma^{u\to v'}_{ij\to k'l'}=\delta s^{u\to v'}-\delta q_{ij\to k'l'}
\end{equation}
where $\delta q_{ij\to k'l'}=-\log(Z^{{\gamma}^{-1}}_{ij} Z^{\gamma_F}_{k'l'})$. The entropy production distribution $\sigma$ is the same as \cref{DEP}. The quasiprobability distribution of the TPM for the backward process can be defined as
\begin{align}
    {P'}^{u, v'}_{ij, k'l'}= p'_{v'}   (\Phi^{AS}\otimes \Pi_0^{S'}|  \mathcal{R}_{SS'}^\gamma|{\Pi'}_{ij}^A\otimes{\Pi'}_{k'l'}^{SS'}) \notag \\
      \times ({ \Pi'}_{ij}^A\otimes{\Pi'}_{k'l'}^{SS'}  |\Pi_{\psi_u}^A\otimes\Pi_{\phi'_{v'}}^{SS'}).
\end{align}
The quasiprobability distribution for the backward process also satisfies the marginality condition:
\begin{align}
    \sum_{u,w'',ij, k'l'}   {P'}^{u, v'}_{ij, k'l'}=  p'_{v'}  (I^{S}\otimes \Pi_0^{S'}|   \mathcal{R}_{SS'}^\gamma  | \Pi_{\phi'_{v'}}^{SS'})  =   p'_{v'},\notag \\
    \sum_{v',ij, k'l'}   {P'}^{u, v'}_{ij, k'l'}=  (\Pi_{\psi_u}^S\otimes \Pi_0^{S'}|   \mathcal{R}_{SS'}^\gamma  | \rho_F^{SS'})  =:   p'_{u},
\end{align}
where we use the property that the final states of $\mathcal{R}_{SS'}^\gamma$ are always in the form of $\rho^S\otimes {\Pi_0}^{S'}$. The relation in \cref{FTR} still holds. Combining \cref{COEPTcps,SIGOTPcps}, we obtain
\begin{equation}\label{TPSIGtr}
    \braket{\sigma}=S(\rho_I||\gamma)-S(\rho^{SS'}_F||\gamma^{SS'}_F).
\end{equation}
 Comparing this result with \cref{NMTRS}, we find that the average entropy production in \cref{TPSIGtr} contains partial information about the intermediate state $\Tr_{S}\rho^{SS'}_F=\sum_{k}  \Pi_{k} \rho_M \Pi_{k}^\dagger$. Compared with that provided by \cref{MTRS}, the information here is not complete. Similar to \cref{HNMEFT}, the entropy production here can also be related to the degree of non-Markovianity. Unlike the Markov process $\mathcal{N}^{\text{Markov}}=\mathcal{N}_2\circ\mathcal{N}_1$ used in \cref{BLPM}, the process $\mathcal{N}^{\text{Markov}}_{SS'}=\mathcal{N}_S^2\circ\mathcal{A}_{ SS'} \circ \mathcal{N}_S^1$ that is closest to $\mathcal{N}_{SS'}$. \cref{MMBC} is used here as a sufficient condition for $\mathcal{N}_{SS'}=\mathcal{N}^{\text{Markov}}_{SS'}$. This means that the quasidistance between $\mathcal{N}_{SS'}$ and $\mathcal{N}^{\text{Markov}}_{SS'}$ can measure the degree of non-Markovianity. To the best of our knowledge, such a non-Markovianity measure has not yet been discussed. We briefly discuss it in \cref{ANNMM}. Further research will be needed to better understand this non-Markovianity measure.

The mapping $\mathcal{A}^i$ does not change the quantum relation entropy, so we can rewrite \cref{TPSIGtr} as
\begin{align}\label{EFUMA}
    \braket{\sigma}= S(\rho_I^S||\gamma^S)- S(\rho_M^{S}||\gamma^{S}_M)\notag\\
    +S(\rho_M^{SS'}||\gamma^{SS'}_M)-S(\rho_F^{SS'}||\gamma^{SS'}_F),
\end{align}
where $\rho_M^{SS'}=\mathcal{A}_{ SS'}(\rho_M^S\otimes \Pi_0^{S'})$ and $\rho_M^{S}=\mathcal{N}_S^1 (\rho_I^S)$. 
The quantity $S(\rho_I^S||\gamma^S)- S(\rho_M^{S}||\gamma^{S}_M)$ is nonnegative. When setting $\mathcal{N}_1=\Tr_E \mathcal{U}_{SE}^{1}\circ\mathcal{A}$, this value matches the entropy production of the first step. The quantity $S(\rho_M^{SS'}||\gamma^{SS'}_M)-S(\rho_F^{SS'}||\gamma^{SS'}_F)$ is nonnegative when no memory effect is present. For a non-Markovian process, we can separate the contributions as follows:
\begin{align}\label{DCONM}
    \braket{\sigma}=S(\rho_I^S||\gamma^S)-S(\rho_M^{S}||\gamma^{S}_M)+S(\rho_M^{SS'}||\gamma^{SS'}_M)\notag \\
    -S(\mathcal{N}_2^S(\rho_M^{SS'})||\mathcal{N}_2^S(\gamma^{SS'}_M))-\sigma_{NM},
\end{align}
where the nonnegative quantity
 \begin{equation}\label{NMM}
    \sigma_{NM}=S(\rho_F^{SS'}||\gamma^{SS'}_F)-S(\mathcal{N}_2^S(\rho_M^{SS'})||\mathcal{N}_2^S(\gamma^{SS'}_M)))
 \end{equation}
 is similar to \cref{BLPM}. This quantity also reflects memory effects and can reduce system fluctuations.

\section{Conclusion and outlook}
In this paper, we first discuss the relation between the FTs for closed quantum systems and the FTs for quantum channels. We find that the FTs are equivalent when utilizing a special map to determine the initial state of the backward process. After that, we extend the FTs for quantum channels to multitime processes. We use a many-body channel and its derived channel to provide a general framework for multitime processes. For Markovian processes, we show that the two-point measurements can be extended to a multipoint measurement. We prove the corresponding FTs and find that the total fluctuation is the aggregation of the fluctuations of each step. For non-Markovian processes, we find that the multipoint measurement yields an ill-defined quasiprobability distribution. The complete measurements of the intermediate states lead to conflicts. Then, we insert operations between the steps and use the derived channel to obtain the corresponding FTs. The inserted operations convey partial information about the intermediate states. The given FTs show that memory effects can reduce the system fluctuations.

The transition matrices are formed by the inner product of the measurement operator and the Choi state, and it would be interesting to research the FTs based on the Choi state. According to the Choi-Jamio\l{}kowski isomorphism, a trace preserving map is completely positive if and only if its Choi state is nonnegative $\mathcal{N}^{S}|\Phi^{AS})\geq 0$. The positivity of the Choi state may be the key to this approach.  

The non-Markovianity measure in \cref{ANNMM} is related to the general measurements of the intermediate system. It is crucial to understand how memory effects influence fluctuations. It would be interesting to find the deeper physical meaning of $\sigma_{NM}$. Comparing it with other non-Markovianity measures might help to find the answer to this question.

The average entropy production depends on the initial system state. In \cref{HOLI}, we find that this quantity is not a linear function of the density matrix. The induced change in Holevo information must be accounted for. It may be useful to understand how the initial state of a system impacts its entropy production.

\begin{acknowledgments}
   ZH is supported by the National Natural Science Foundation of China under grant nos. 12047556, 11725524 and the Hubei Provincial Natural Science Foundation of China under Grant No. 2019CFA003.
\end{acknowledgments}

\appendix
\section{Non-Markovianity measure}\label{ANNMM}
For a general many-body channel $\mathcal{N}^{S^1S^2}$, the operation $ \mathcal{A}^i$ in \cref{NMCQPT} yields 
\begin{equation}\label{AGMTMBC}
    \sum_{kl}(\Pi_{kl}^{S_1}|\mathcal{N}^{S^1S^2}|\Pi_{kl}^{S_2}\otimes \Pi_{kl}^{S'_1}),
\end{equation}
which maps the initial state $\rho_I^{S_1}$ to the final state $\rho_F^{S_2S'_1}$ ($\rho_I^{S}$ to $\rho_F^{SS'}$ in the main text). For a Markovian process, $\mathcal{N}^{S^1S^2}=\mathcal{N}^{S^2}\circ\mathcal{N}^{S^1}$. And \cref{AGMTMBC} can be expressed as
\begin{equation}
    \sum_{kl}   \mathcal{N}^{S^2} |\Pi_{kl}^{S_2}\otimes \Pi_{kl}^{S'_1})(\Pi_{kl}^{S_1}|\mathcal{N}^{S^1},
\end{equation}
which corresponds to $\mathcal{N}_S^2\circ\mathcal{A}_{ SS'} \circ \mathcal{N}_S^1$ in the main text. Hence, \cref{MMBC} is a sufficient condition for obtaining $\mathcal{N}_{SS'}=\mathcal{N}^{\text{Markov}}_{SS'}$.

When $\mathcal{N}_{SS'}=\mathcal{N}^{\text{Markov}}_{SS'}$, the $\sigma_{NM}$ in \cref{NMM} is equal to zero. Only memory effects can lead to an increase in distinguishability and allow $S(\rho_F^{SS'}||\gamma^{SS'}_F)>S(\rho_M^{SS'}||\gamma^{SS'}_M)$. 

Recall that in \cite{PRFPM18}, the quasidistance between the generalized Choi state of a non-Markovian process and the closest Choi state of a Markov process measures the degree of non-Markovianity:
\begin{equation}
    \mathcal{D}_{NM}:=\min_{\Upsilon^{\text{Markov}}}\mathcal {D}(\Upsilon||\Upsilon^{\text{Markov}}),
\end{equation}
where $\Upsilon$ is the Choi state of a many-body channel $ \mathcal{N}^{S^1\dots S^n}$ and $\Upsilon^{\text{Markov}}$ is the Choi state of a Markov process $ \mathcal{N}^{S^n}_n \circ \dots \circ  \mathcal{N}^{S^1}_1$. This non-Markovianity measure is also suitable for the derived channels in \cref{EQC}. The Markovian process yields
\begin{equation}
    \mathcal{N}^{\text{Markov}}_{\textbf{A}_{n-1:1}}= \mathcal{N}_n\circ(\mathcal{A}_{n-1} \circ \mathcal{N}_{n-1})\circ\ldots \ldots\circ(\mathcal{A}_1\circ \mathcal{N}_1).
\end{equation}
Any CP-inducing quasidistance between the Choi state of $  \mathcal{N}^{S}_{\textbf{A}_{n-1:1}}$ and the closest Choi state of $ \mathcal{N}^{\text{Markov}}_{\textbf{A}_{n-1:1}}$ also measures the degree of non-Markovianity.

\section{Holevo information}\label{HOLI}
The entropy production in \cref{FOAV} is state-dependent. Suppose that the initial state $\rho=\sum_a p_a \rho_a$ yields $ \braket{\sigma}$ and the initial states $\rho_a$ produce $ \braket{\sigma_a}$; then, it is easy to show that
\begin{equation}
    \braket{\sigma}= \sum_a p_a \braket{\sigma_a}+\delta \chi,
\end{equation}
where $ \delta \chi= \chi_I- \chi_F$. $ \chi_I$ is the Holevo information of $\rho$, and $ \chi_F$ is the Holevo information of $\mathcal{N}(\rho)$.

\end{CJK*}

\end{document}